\documentclass{amsart}

\usepackage{amsmath,amscd,amssymb,amsthm}
\usepackage{latexsym,enumerate,graphicx,geometry,rotating,braket,mathrsfs,url}

\newcommand{\bibfilename}{unibib}

\numberwithin{equation}{section}

\theoremstyle{plain}
\newtheorem{thm_}[equation]{Theorem}
\newtheorem{lemma_}[equation]{Lemma}
\newtheorem{prop_}[equation]{Proposition}
\newtheorem{cor_}[equation]{Corollary}
\newtheorem{eg_}[equation]{Example}
\newtheorem{con_}[equation]{Conjecture}
\newtheorem*{cons_}{Conjecture}

\theoremstyle{definition}
\newtheorem{thmu_}[equation]{Theorem}
\newtheorem*{thmus_}{Theorem}
\newtheorem{propu_}[equation]{Proposition}
\newtheorem*{propus_}{Proposition}
\newtheorem{coru_}[equation]{Corollary}
\newtheorem{lemu_}[equation]{Lemma}
\newtheorem{egu_}[equation]{Example}
\newtheorem*{egus_}{Example}
\newtheorem{def_}[equation]{Definition}
\newtheorem*{defs_}{Definition}

\theoremstyle{remark}
\newtheorem{rk_}[equation]{Remark}

\newcommand{\thm}[1]{\begin{thm_}#1\end{thm_}}
\newcommand{\thmu}[1]{\begin{thmu_}#1\end{thmu_}}

\newcommand{\lemm}[1]{\begin{lemma_}#1\end{lemma_}}
\newcommand{\lemu}[1]{\begin{lemu_}#1\end{lemu_}}

\newcommand{\prop}[1]{\begin{prop_}#1\end{prop_}}

\newcommand{\defi}[1]{\begin{def_}#1\end{def_}}

\newcommand{\rk}[1]{\begin{rk_}#1\end{rk_}}

\newcommand{\con}[1]{\begin{con_}#1\end{con_}}

\newcommand{\pf}[1]{\begin{proof}#1\end{proof}}
\DeclareMathOperator{\ord}{ord}
\DeclareMathOperator{\GL}{GL}
\DeclareMathOperator{\Gal}{Gal}

\newcommand{\fracn}[2]{\genfrac{(}{)}{}{}{#1}{#2}}

\newcommand{\ZZ}{\mathbb Z}
\newcommand{\QQ}{\mathbb Q}
\newcommand{\RR}{\mathbb R}
\newcommand{\CC}{\mathbb C}

\newcommand{\FF}{\mathbb F}

\renewcommand{\a}{\mathfrak a}
\newcommand{\vA}{\mathscr A}

\renewcommand{\o}{\mathfrak o}
\newcommand{\p}{\mathfrak p}
\renewcommand{\P}{\mathfrak P}
\newcommand{\vP}{\mathscr P}

\newcommand{\SSS}{\mathbf S}
\renewcommand{\t}{\times}

\newcommand{\lr}{\longrightarrow}
\newcommand{\llr}{\longleftrightarrow}

\newcommand{\hr}{\hookrightarrow}

\newcommand{\s}{\sigma}

\newcommand{\eq}[1]{\begin{equation}#1\end{equation}}
\newcommand{\eqn}[1]{\begin{equation*}#1\end{equation*}}
\newcommand{\ga}[1]{\begin{gather}#1\end{gather}}
\newcommand{\gan}[1]{\begin{gather*}#1\end{gather*}}

\newcommand{\aln}[1]{\begin{align*}#1\end{align*}}

\newcommand{\cs}[1]{\begin{cases}#1\end{cases}}

\newcommand{\itmz}[1]{\begin{itemize}#1\end{itemize}}
\newcommand{\enmt}[1]{\begin{enumerate}#1\end{enumerate}}
\renewcommand{\it}{\item}

\newcommand{\itm}[1]{\it[\upshape{(#1)}]}
\newcommand{\newnoindbf}[1]{\vspace{2mm}\noindent\textbf{#1}}

\begin{document}
\title[Non-existence of GBF]{On the Non-existence of certain classes of generalized bent functions}
\author[C. Lv]{Chang Lv}
\address{
SKLOIS, Institute of Information Engineering, CAS, Beijing 100093, China
}
\email{lvchang@amss.ac.cn}

\author[J. Li]{Jianing Li}
\address{
Key Laboratory of Mathematics Mechanization,
NCMIS, Academy of Mathematics and Systems Science,
CAS, Beijing 100190, China}
\email{lijianing19891026@163.com}


\date{\today}
\begin{abstract}
We obtain  new non-existence results of generalized bent functions from $\ZZ^n_q$ to $\ZZ_q$ (called type $[n,q]$).
The first case is a class of types where $q=2p_1^{r_1}p_2^{r_2}$.  
The second case contains two types 
$[1\le n\le3,2\t31^e]$ and
$[1\le n\le7,2\t151^e]$.
\end{abstract}
\maketitle

\section{Introduction}\label{sec_intro}
Let $q\geq 2$ be an integer, $\ZZ_q=\ZZ/{q\ZZ}$, $\zeta_q=\textnormal{exp}(\frac{2\pi i}{q})$.
\defi{\label{def_gbf}
 A function $f$ from $\ZZ^n_q$ to $\ZZ_q$ 
is called a \emph{Generalized Bent Function} (\emph{GBF}) of \emph{type} $[n,q]$ if
\eq{\label{eq_F_abs}
 F(\lambda)\overline{F(\lambda)}=q^n 
}
 for every $\lambda\in \ZZ^n_q$ where
\eqn{
 F (\lambda)=\sum_{x\in\ZZ^n_q}\zeta^{f(x)}_q\cdot \zeta^{-x\cdot \lambda}_q 
}
 is the Fourier transform of the function $\zeta^{f(x)}_q$,
 $x\cdot \lambda$ is the standard dot product, 
and $\overline{F(\lambda)}$ is the complex conjugate of $F(\lambda)$.
}

Bent functions were first introduced by Rothaus \cite{Rothaus} in 1976,
and  were generalized to GBFs  by Kumar et al. \cite{Kumar} in 1985.
GBFs have been used in many fields such as  difference sets, coding theory, cryptography and sequence designs.
For more background information and its applications we refer the reader to \cite{Dillon,OSW,Kumar}.

A natural question is when bent functions do exist. 
Rothaus \cite{Rothaus} proved that bent functions from $\ZZ_2^n$ to $\ZZ_2$ exist if and only if $n$ is even. 
For GBFs  defined in \ref{def_gbf}, Kumar et al. \cite{Kumar} constructed them except the case that $n$ is odd and $q\equiv2\pmod 4$.

From now on we assume $n$ is odd and $q=2N$ with $2\nmid N\ge3$ 
So far there is no GBF being constructed in this case,
while there are many  non-existence results  under some extra constraints.
We give a list of these results with reference at the beginning of each item:
\enmt{
\it (Kummar \cite{Kumar})  $2^s\equiv-1\pmod N$ for some  integer $s\ge1$,
\it (Ikeda \cite{Ikeda})\label{it_ikeda}  type $[1, 2p_1^{e_1}\dots p_g^{e^g}]$ where $p_1,\dots,p_g$ are distinct primes\\
 and $p_i^{s_i}\equiv-1\pmod{N/p_i^{e_i}}$ for some $s_i, i=1,\dots,g$.
}

By \eqref{eq_F_abs} in the definition, 
if there is no element in $\ZZ[\zeta_q]$ with absolute value $q^{\frac{n}{2}}$,
i.e. 
\eq{\label{eq_diophantine}
\alpha\bar\alpha=q^n,
} 
has no solution in $\ZZ[\zeta_q]$.
 then there is no GBF of type $[n,q]$. 
Using this fact,  Feng et al. \cite{feng-gbf1,feng-gbf2,Feng3} obtain the following non-existence results:
\enmt{
\setcounter{enumi}{2}
\it (Feng \cite{feng-gbf1})\label{it_feng1} type $[n<m/s, 2p^l]$, where $p\equiv7 \pmod8$ is a prime, $s=\frac{\varphi(p^l)}{\ord_{p^l}(2)}$  and $m$ is the smallest odd positive integer s.t. $x^2+py^2=2^{m+2}$ has integral solutions,
\it (Feng et al. \cite{feng-gbf1,feng-gbf2,Feng3,Liu-Feng}) various classes of type $[n<m,2p_1^{l_1}p_2^{l_2}]$, where $p_1,p_2$ are two distinct primes satisfy some conditions and $m$ is an upper bound for $n$.
} 

 However,   in the case that
\eqref{eq_diophantine} is  solvable over $\ZZ[\zeta_q]$,
one has  to  use other methods to obtain non-existence results,
which are typically collected in the following: 
\enmt{ 
\setcounter{enumi}{4}
\it (Pei \cite{Pei}) type $[1,2\times 7]$,
\it (Ikeda \cite{Ikeda})  see \eqref{it_ikeda} before,
\it (Jiang and Deng \cite{jiang-bent}) $[3,2\times 23^e]$,
\it (Li and Deng \cite{li-bent})\label{it_li}  type $[m,2p^e]$ where  $p\equiv7 \pmod 8$ is a prime with $\ord_{p^e}(2)=\varphi(p^e)/2$
and $m$ is defined the same as in \eqref{it_feng1}.
}

In this article we extend the results where $q=2p^e$ in \eqref{it_li}  to the case where  $q=p_1^{r_1}p_2^{r_2}$:
\thm{ \label{thm_pq} 
Let  $N=p_1^{r_1}p_2^{r_2}$ where $r_1,r_2\ge1$. Let  $p_1\equiv7 \pmod 8$  and $p_2\equiv 5\pmod 8$ be two primes
satisfy \enmt{
\itm{i} $\ord_N(2) = \varphi(N)/2$, where $\varphi$ is the Euler phi function;
\itm{ii} there is a positive integer $s$ such that $p_1^s\equiv-1\pmod{p_2^{r_2}}$;
\itm{iii} there is a positive integer $t$ such that $p_2^t\equiv-1\pmod{p_1^{r_1}}$.
}
Let $m$ be the smallest odd positive integer such that $x^2+py^2=2^{m+2}$ has integral solution $(x,y)$.
Then there is no GBFs of  type $[m,2N]$.
}
The above case is different from the ones in Feng et al. \cite{feng-gbf1,feng-gbf2,Feng3,Liu-Feng} 
so the result is new. See Remark \ref{thm_pq.rk_new} for details.

And we also  obtain  the non-existence results of  two types 
$[1\le n\le3,2\t31^e]$ and
$[1\le n\le7,2\t151^e]$:
\thm{ \label{thm_sp} 
Let $e$ and $n$ be  positive integers and $n$ be  odd.
Then  there is no GBFs of  type
$[1\le n\le3,2\t31^e]$ and
$[1\le n\le7,2\t151^e]$.
}
\rk{
The result in  \cite[Theorem $3.1$]{feng-gbf1} can only deal with  types $[n<m/s, p^l]$ where
$s=\ord_{p^l}(2)$ and $p\equiv7\pmod8$. By the calculation in \cite[pp. 566]{feng-gbf1},
$m/s=1,7/5$ for $p=31,151$. Thus the results in the above theorem is also new.
}
 
For the proofs of the two theorems, we need some facts in algebraic number theory which are contained 
in Section \ref{sec_facts}.
And for the cases where
\eqref{eq_diophantine} is  solvable over $\ZZ[\zeta_q]$,
we generalize the idea used in \cite{Pei, Ikeda, jiang-bent, li-bent}, which we 
name as the   \emph{element partition method} and prove in Section \ref{sec_epm}. 
With these preparations, we can prove Theorem \ref{thm_pq} and \ref{thm_sp}
in Section \ref{sec_pq} and \ref{sec_sp}, respectively.
Finally, we give some additional non-existence results obtained by \url{PARI/GP} \cite{pari} without proofs,
and some remarks concerning the future work (Section \ref{sec_pari}).

\section{Basic facts of algebraic number theory}\label{sec_facts}
 The methods of proving non-existence results of GBF are often involve algebraic number theory,
  mainly the basic arithmetic (ideals, units, class groups etc.) of cyclotimic fields and their subfields.  The standard reference is \cite[Chapter~$2,3$]{Marcus}  or \cite[Chapter~$2$]{Washington}. In this section, we list some facts needed in the proof.

For any number field  $F$, denote by $\o_F$ the ring of integers of $F$,
by $Cl(F)$ the  class group of $F$ and by $h(F)$ the class number of $F$.

Let $h_N=h(\QQ(\zeta_N))$
and $h_N^+=h(\QQ(\zeta_N+\zeta_N^{-1}))$. It is well-known that
$h_N^+\mid h_N$ and thus one can write $h_N=h_N^+h_N^-$.

For the subfield of the cyclotomic field $\QQ(\zeta_{p^e})$ where $p^e$ is a prime power,
we have the divisibility of class numbers.
\lemm{\label{lem_h_div}
Let  $L=\QQ(\zeta_{p^e})$.
Let $F$ be a subfield of $L$. Then we have $h(F)\mid h(L)$.
}
\pf{
Let $H_F$ (resp. $H_L$) be the Hilbert class field of $F$ (resp. $L$).
Then $H_F/F$ is unramified abelian, so the same is true for $LH_F/L$.
It follows that $LH_F\subseteq H_L$. But $H_F\cap L=F$ in our case since
$p$ is totally ramified in $L$. Hence 
\eqn{
h(F)=[H_F:F]=[LH_F:L]\mid [H_L:L]=h(L).
}
}

For more specific cases, we have the  more strong
\prop{\label{prop_cl_inj}
Let $p\equiv3\pmod4$ and $L=\QQ(\zeta_{p^e}), F:=\QQ(\sqrt{-p})$.
It is well-known that  $F$ is a subfield of $L$.
Let $E$ be any number field such that $F\subseteq E\subseteq L$.
Then the canonical morphism $j_{E/F}: Cl(F)\lr Cl(E)$ sending 
$\a$ to $\a\o_E$ is injective.
}
\pf{
Let $I_F$ be the group of fractional ideals of $F$, and $P_F$ be principal ones.
Let $I_E$ and $P_E$ be the corresponding groups for $E$.
We know the Galois group $G:=\Gal(E/F)$ acts on $P_E$  and $\o_E^\t$.
Then by Greenberg \cite[Proposition~$1.2.3$]{greenberg2006topics} we have
\eq{\label{prop_cl_inj.eq_h1}
\ker(j_{E/F})\hr P_E^G/P_F\cong H^1(G,\o_F^\t).
}
Let $n=[E:F]$ then $n\mid \varphi(p^e)/2$ is odd since $p\equiv3\pmod4$.
Thus \eqn{
\o_F^\t=\{\pm1\}=N_{E/F}(\pm1)\subseteq N_{E/F}(\o_E^\t).
}
Note $E/F$ is cyclic. Thus we have  $H^2(G,\o_E^\t) = H^0(G,\o_E^\t)$ is trivial.

On the other hand, Greenberg \cite[Proposition~$1.2.4$]{greenberg2006topics} tells us that 
\eqn{
\#H^1(G,\o_E^\t)=n\#H^2(G,\o_E^\t).
}
 Combining with \eqref{prop_cl_inj.eq_h1} we know 
\eqn{
\#(P_E^G/P_F)=n.
}
Let $\P$ be a prime of $E$ dividing $p$ and $\p=\P\cap F$.
Since $p$ is totally ramified in $L\supseteq E$ we know   $\P\in P_E^G/P_F$.

Now we claim that $\P$  has order $n$ and hence $P_E^G/P_F$ is generated by $\P$.
Actually, $\P^n=\p\o_E\in P_F$. On the other hand, if we have $\P^k\in P_F$ for 
some $k$, then $\P^k=\alpha\o_E$ for some $\alpha\in F$. It follows that
\eqn{
\ord_{\p}(\alpha)n=\ord_{\P}(\alpha)=k.
}
Hence $n\mod k$. This proves the claim  and 
that $P_E^G/P_F$ is generated by $\P$.

Now if $\a\in\ker(j_{E/F})$ then $\a\o_E=\beta\o_E$ for some $\beta\in E$.
Consider its image in  $P_E^G/P_F=\left<\P\right>$ so we have $\a\o_E=\P^l\gamma$ for some $l$ and 
$\gamma\in F$. Again the fact that 
\eqn{
\ord_{\p}(\a/\gamma)n=\ord_{\P}((\a/\gamma)\o_E)=l
}
 gives $n\mid l$, which is to say $\a$ is trivial in $P_E^G/P_F$.
Then the injection in \eqref{prop_cl_inj.eq_h1} implies that $\ker(j_{E/F})$ is trivial
and we complete the proof of the proposition.
}

Next we introduce  Stickelberger ideals.
Suppose $p$ is  a prime and $K_0=\QQ(\zeta_p)$ 
and $G_0=\Gal(K_0/\QQ)\cong(\ZZ/p\ZZ)^\t$.
\defi{
The \emph{Stickelberger element} $\theta=\theta_p\in\QQ[G_0]$ 
is defined by \eqn{
\theta=\sum{a\in(\ZZ/p\ZZ)^\t}\left\{\frac{a}{p}\right\}\s_a^{-1}
} where $\{\frac{a}{p}\} = \frac{a}{p}- p[\frac{a}{p}]$.
And the \emph{Stickelberger ideal} $S_p$ of $\ZZ[G_0]$ is defined by \eqn{
S_p=\ZZ[G_0]\theta\cap\ZZ[G_0].
}
}
We mainly use these following properties of Stickelberger ideal:
\prop{\label{prop_stick}
We have \itmz{
\itm{a} For $(c, p) = 1$, the element $(c-\s_c)\theta$ are in $S_p$.
\itm{b} The Stickelberger ideal $S_p$ annihilates the ideal class group $Cl(M)$,
where $M$ is a subfield of $K_0$ such that  $p$ is the minimal integer with the property that
$M\subseteq \QQ(\zeta_p)$.
}}
\pf{ See \cite[Lemma $6.9$ and Theorem $6.10$]{Washington}.}

Through this  paper, we fix the following notations.
Let $q=2N$ with $2\nmid N\ge3$,  $\zeta=\zeta_N$ be an $N$-th primitive root of unity,
 $K=\QQ(\zeta)$, and $D\subset K$ be the decomposition field of $2$ in $K$.
Let $G=\Gal(K/\QQ)$. It's well-known that $G\cong(\ZZ/N\ZZ)^\t$, the isomorphism being 
$c\mapsto(\s_c: \zeta\mapsto \zeta^c)$ for $\gcd(c, N)=1$.

  For our purpose we need to investigate the equation $\alpha\bar{\alpha}=q^n=(2N)^n$ where $\alpha \in \ZZ[\zeta_N]$. So we first study the idealic behaviour of $2$ and $p$ in the cyclotomic field $K$.
 
The following lemma taken from Feng \cite[Lemma $2.1$]{feng-gbf1} will \emph{descent} the equation $\alpha\bar{\alpha}=2^n$, $\alpha \in \o_K$ to $\alpha\bar{\alpha}=2^n$,  $\alpha \in \o_E$, where  $E$ is a subfield of $K$.
\lemu{[\cite{feng-gbf1}, Lemma $2.1$]\label{lem_descent} 
 	If $\alpha \overline{\alpha}=2^n$ for some $\alpha\in \o_K$ and a positive integer $n$, then there exist 
$\beta=\pm\zeta^j\alpha$ for some $j\in\ZZ$ 
and subfield $E\subseteq K$ containing $D$ with  $[E:D]\le2$
such that
 $\beta\in\o_E$ and $\beta\bar\beta=2^n$. 
}

\section{The element partition method}\label{sec_epm}
In this section, we will prove the following
\prop{\label{prop_epm}
Suppose $t$ is an odd positive integer and $q=2N,\ 2\nmid N\ge3$.
Let  $f: \ZZ_q^t\lr \ZZ_q$ be a function with $F(\lambda)$  its  Fourier   transform defined as
before. 
Suppose $F(\lambda)$ has the property that for every $\lambda \in \ZZ ^t _q$ and  $v\in \ZZ^t _q$  an element of order 2, 
\ga{
F(\lambda)\not\subseteq2\o_K \label{eq_notin2}\\
\text{and }F(\lambda)\o_K= F(\lambda+v)\o_K. \label{eq_same_class}
}
Then $f$ is not a GBF.
}
The idea behind this proposition dates back to 
 the  method developed by  Ikeda \cite{Ikeda}  and   Jiang-Deng \cite{jiang-bent}.
To prove the  proposition  we  have to generalize this.

Now we assume $f$ is a GBF so by definition we have $F(\lambda)\overline{F(\lambda)}=q^n$, which leads to the 
following 
\lemm{\label{lem_pm}
Let $\lambda,v, f, F$ be as in Proposition~$\ref{prop_epm}$.
If $f$ is a GBF, we have $F(\lambda)=\pm F(\lambda+v)$.
}
\pf{
Since  $v$ is of order $2$,  we have $\zeta^{x\cdot v}_q=\pm1$. So 
\eq{\label{eq_F+F}
F(\lambda)+F(\lambda+v)
=\sum_{x\in \ZZ^t_q}{\zeta_q ^{f(x)-x\cdot \lambda}(1+\zeta_q ^{-x\cdot v})}
\in 2\o_K.
}
From \eqref{eq_same_class} we know that
$u:=F(\lambda)/F(\lambda+v)\in\o_K^\t$.
For each $\s\in G$, \eqn{
\s(u)\overline{\s(u)}=\s(u\bar u)
=\s\left(\frac{F(\lambda)\overline{F(\lambda)}}{F(\lambda+v)\overline{F(\lambda+v)}}\right)
=1,\quad\text{ (by \eqref{eq_F_abs}) }
}
whence $u$ is a root of unity and so $u=\pm\zeta^j$. 
Thus  \eqn{ F(\lambda)=\pm\zeta^jF(\lambda+v). }
Then \eqn{
F(\lambda)+F(\lambda+v)=F(\lambda)(1\pm\zeta^j).
}
It follow from \eqref{eq_F+F} that $2\mid  F(\lambda)(1\pm\zeta^j)$ in $\o_K$.
 Note that   if $\zeta^j\neq1$ then $1\pm\zeta^j$ is an unit or divides $N$,
so  $2\o_K+(1\pm \zeta^j)\o_K=\o_K$. Thus $2\mid F(\lambda)$,
which  is a contradiction to the hypothesis  \eqref{eq_notin2}.
This shows that  $F(\lambda)=\pm F(\lambda+v)$.
}

Now let us  pin down  some notations.
Let $P_2$ be the Sylow-$2$ subgroup of $\ZZ ^t_q$. Then $P_2\cong \FF^t_2$ as $\FF_2$-vector space and we 
write $P_2=\{0,v_1,\dots,v_{2^t-1}\}$.
   For every $v_i\in P_2$,  we know that $v_i$ has order $2$. If we define  
\aln{
&N_i=N_{v_i}=\set{x\in \ZZ^t_q| F(x)=F(x+v_i)},\\
\text{and }&M_i=M_{v_i}:=\set{x\in \ZZ^t_q| F(x)=-F(x+v_i)},
}
then $\ZZ^t_q=N_i\sqcup M_i$ by Lemma~\eqref{lem_pm}, where the symbol $\sqcup$ means disjoint union.
\rk{
For sake of the above decomposition $\ZZ^t_q=N_i\sqcup M_i$, we call this method element partition.
}
Let $n_i=\#N_i, m_i=\#M_i$ be the cardinality of $N_i,M_i$, so $n_i+m_i=q^t$.
\lemm{ \label{n=m}
	$n_i=m_i=\frac{q^t}{2}$.
}
\pf{
	As $F$ is a Fourier transform of $\zeta^{f(x)}$, which is a function from $\ZZ^t_q$ to $\SSS^1=\set{z\in \CC| |z|=1 }$, we have $$\sum_{x\in \ZZ^t_q}{F(x)\overline{F(x+v_i)}}=0.$$
Then by the definition \eqref{eq_F_abs},   $F(x)\overline{F(x)}=q^t$ for each $x$.
It follows that
	$$0=\sum_{x\in N_i}{F(x)\overline{F(x+v_i)}}+\sum_{x\in M_i}{F(x)\overline{F(x+v_i)}}=q^t(n_i-m_i). $$ Hence, $n_i=m_i$ for each $i$.
}

Here we review the  proof of  Ikeda \cite{Ikeda} for the case $t=1$, since it is the basic idea of this method. 
   First note that  $x\in N_v$ if and only if $x+v\in N_v$, where $v$ is the unique element of order $2$ in $\ZZ_q$.
Hence $2\mid n_v$. But $n_v=\frac{q}{2}$ is odd, which is a contradiction. 

It can be seen that Ikeda \cite{Ikeda} used one $2$-order element to prove the result in  the case $t=1$.  In Jiang-Deng \cite{jiang-bent}, they use three $2$-order elements to treat the case $t=3$. 

We now  generalize this method systematically to treat the general case. The second author used the same augment in \cite{li-bent}
 to prove that there is no GBF of  type $[m,2p^e]$ where  $p\equiv7 \pmod 8$ is a prime with $\ord_{p^e}(2)=\varphi(p^e/2)$
and $m$ is defined the same as in Theorem \ref{thm_pq}.

In the remain of the proof, we assume  $t\geq 3$ so that $\#P_2\geq8$. Now we define $2^{2^t -1}$ subsets of $\ZZ^t_q$ by using all $2$-order elements as follows:
$$X_1\cap X_2\cap \dots \cap X_{2^t-1},$$ with each $X_i=N_i \text{ or }M_i$.
Obviously, $\ZZ^t_q$ is a disjoint union of all these subsets. Our main task is to compute the cardinality of each subsets. 

\lemm{\label{lem_3setcap}
	If $u,v,w\in P_2-\{0\}$ are pairwise different and $u+v+w=0$, then we have 
\eqn{
N_u\cap N_v\cap M_w=M_u\cap M_v\cap M_w=\emptyset.
}
}
\pf{
We only need the assumption  \eqref{eq_notin2} that  $F(\lambda)\notin 2\o_K$,
 so the proof below is essentially same as the case $p=23$ as in 
\cite[Lemma~$11$]{jiang-bent}.
	
	First, note that \aln{
x&\in N_u\cap N_v\cap M_w\\
\iff x+w &\in M_u\cap M_v\cap M_w.
}
So it's enough to prove $M_u\cap M_v\cap M_w=\emptyset$.

Second, note that the map 
\eqn{
 y\mapsto y\cdot u: \ZZ^t_q \lr \{0,\frac{q}{2}\}\subset\ZZ_q
}
is surjective, so 
\eqn{
 \zeta^{y\cdot u}=\cs{
 1&\text{ if }y\cdot u=0\\
-1&\text{ if }y\cdot u\not=0.
}
}
	 For simplicity, we write $\sum=\sum{\zeta^{f(y)-x\cdot y}}$.
	 Now take an element $x\in M_u\cap M_v\cap M_w$, then
\aln{
F(x)&=\sum_{y\in \ZZ^t_q}=\sum_{y\cdot u=0}+\sum_{y\cdot u\neq0}\\
&=-F(x+u)=-\sum_{y\cdot u=0}+\sum_{y\cdot u\neq0}.
}
	
	So we get $$0=\sum_{y\cdot u=0}=\sum_{y\cdot u=0, y\cdot v=0	}+\sum_{y\cdot u=0,y\cdot v\neq0}.$$
	
	Similarly, we have $$0=\sum_{y\cdot v=0}=\sum_{y\cdot u=0, y\cdot v=0	}+\sum_{y\cdot u\not=0,y\cdot v=0}.$$
	
	Also since $$0=\sum_{y\cdot w=0},$$ we have that
	$$F(x)=\sum_{y\cdot w\not=0}=\sum_{y\cdot (u+v)\neq0}=\sum_{y\cdot u=0,y\cdot v\neq0}+\sum_{y\cdot u\not=0,y\cdot v=0}=-2\sum_{y\cdot u=0,y\cdot v=0}\in 2\o_K.$$ This contradict to \eqref{eq_notin2}.	 So $M_u\cap M_v\cap M_w=\emptyset$.
	
}

 Then we have the following observation. It tells us among the $2^{2^t-1}$ subsets, there are at most $2^t$ nonempty subsets. And  these $2^t$ subsets are rather ``nice". 

\lemm{ \label{observation}
	Let $N_i,M_i$ be as above, $X_1\cap X_2\cap \dots \cap X_{2^t-1}\subseteq \ZZ^t_q$ with $X_i=N_i$ or $M_i$. 
If $\set{v_i\in P_2|X_i=N_i}\cup\{0\}$ is not a subgroup of $P_2$ with index $1$ or $2$, then it must be empty.
}
\pf{
	If  $A:=\set{v_i\in P_2|X_i=N_i}\cup\{0\}$ is not a subgroup, then there are $u,v\in A-\{0\}$ such that $u+v\notin A$, hence $X_u=N_u$, $X_v=N_v$ and $X_{u+v}=M_{u+v}$. Then by Lemma \ref{lem_3setcap}, 
\eqn{
X_1\cap X_2\cap \dots \cap X_{2^t-1}\subseteq N_u\cap N_v\cap M_{u+v}= \emptyset.
}
 If $A$ is a subgroup with index larger than $2$. Then $A$ is  also a $\FF_2$ vector subspace of $P_2$ and  
its $\FF_2$-dimension is less than or  equal    to ${t-2}$.
Thus the dimension of its  complement subspace  $\bar A$ is greater than or  equal to  $2$, 
so we can take $u,v\in \bar A$ such that they are independent.
Then $u,v,u+v\in \bar A-\{0\}$, so $u,v,u+v$  are not in $A$ and 
 hence $X_u=M_u, X_v=M_v$ and $X_{u+v}=M_{u+v}$. Then by Lemma \ref{lem_3setcap} we
 have
\eqn{
X_1\cap X_2\cap \dots \cap X_{2^t-1}\subseteq M_u\cap M_v\cap M_{u+v}= \emptyset
}
and we finish the proof.
}

The following lemma is a basic fact about the subgroups of $P_2$.
 \lemm{ \label{subgroups of $P_2$}
 	There are $2^t-1$ subgroups of $P_2$　with index $2$. 
 }
\pf{
The correspondence
\gan{
\text{subgroups of index} 2\\
=\{ \FF_2 \text{ vector subspace of $P_2$ with dimension } t-1\}\\
\llr  \{ \FF_2 \text{ vector subspace  with dimension  } 1 \} 
} is one-to-one by taking the complement subspace.
	Since there are $2^t-1$ nonzero elements in $P_2$, there are $2^t-1$ subgroups with index $2$ by the above correspondence.
}
	
Let $\{\text{subgroups of $P_2$ with index 2}\}=\{ H_1,\dots,H_{2^t -1} \}$.	 Among the $2^{2^t-1}$ subsets $X_1\cap X_2\cap \dots \cap X_{2^t-1}$ with $X_i=N_i$ or $M_i$, we only need to consider the following subsets, since others are empty sets by Lemma~\ref{observation}:
\aln{
Y_0 &:= \bigcap_{i=1}^{2^t-1} {N_i} &\text{ (corresponding to the subgroup with index $1$)}\\
\text{ and }Y_i &:= \left(\bigcap_{v\in H_i-\{0\}} N_{v}\right)\bigcap \left(\bigcap_{u\in P_2- H_i} M_u\right), &\\
 &\quad i=1,2,\dots,2^t-1. &\text{ (corresponding to subgroups with index $2$)}
}

Let $y_0=\#Y_0$ and  $y_i=\#Y_i$ for $i=1,2,\dots,2^t-1$.

\pf{[Proof of the Proposition~$\ref{prop_epm}$]
Recall that $\ZZ^t_q$ is a disjoint union of the following $2^{2^t-1}$ subsets  $X_1\cap X_2\cap \dots \cap X_{2^t-1}$ where each $X_i=N_i$ or $M_i$.
 Since the only possibly nonempty subsets are $Y_0,\dots,Y_{2^t-1}$ by Lemma~\ref{observation} and Lemma~\ref{subgroups of $P_2$},
we have  $$\ZZ^t_q=\bigsqcup_{i=0}^{2^t-1}Y_i,$$ 
from which we obtain an equation
\eq{\label{eq_hi+ng}
y_0+\sum_{i=1}^{2^t-1} y_i =q^t.
}

 On the other hand 
\aln{
N_i&=N_i\cap \ZZ^t_q=N_i\bigcap\left(\bigsqcup_{i=0}^{2^t-1} Y_i\right)\\
&=Y_0\bigcup\left(\bigsqcup_{1\le j\le 2^t-1,\ v_i\in H_j} Y_j\right), \quad i=1,\dots,2^t-1.
}
Note that $n_i=\frac{q_t}{2}$  by  Lemma~\ref{n=m}. We obtain 
\eqn{
\frac{q^t}{2} = n_i = y_0 + \sum_{1\le j\le 2^t-1,\ v_i\in H_j} y_j,\quad i=1,\dots,2^t-1.
}
By summing up the above  $2^{t}-1$ equations, we obtain 
\aln{
(2^t-1) \frac{q^t}{2} &= \sum_{i=1}^{2^t-1} \left( y_0 + \sum_{1\le j\le 2^t-1,\ v_i\in H_j} y_j \right)\\
   &= (2^t-1) y_0 + \sum_{j=1}^{2^t-1} y_j  \sum_{1\le i\le 2^t-1,\ v_i\in H_j} 1\\
   &= (2^t-1) y_0 + \sum_{j=1}^{2^t-1} y_j  \#(H_j-\{0\}) \\
   &= (2^t-1) y_0 + (2^{t-1}-1) \sum_{i=1}^{2^t-1} y_i. \quad\text{ (since $H_j\subset P_2$ with index $2$) }
}
Conbining   with \eqref{eq_hi+ng}, we treat them as
  two linear equations with two unknown variables $y_0$ and $\sum_{i=1}^{2^t-1} y_i$.
Soving  the equation we have  $y_0=\frac{q^t}{2^t}=N^t$ is an odd number.

However, fixing any $v\in P_2-\{0\}$,  we have $x\in Y_0$ if and only if 
 $x+v\in Y_0$, so $2\mid y_0$. This contradiction shows that 
$f$ is not a GBF.
}

\section{Non-existence result for GBFs of the type $[m, 2p_1^{r_1}p_2^{r_2}]$}\label{sec_pq}
In this section, we will prove  Theorem \ref{thm_pq}, where 
 $N=p_1^{r_1}p_2^{r_2}$ satisfy $($i$)$, $($ii$)$ and $($iii$)$.
Assume $f$ is a GBF of type  $[n=m, q=2N]$. Then $F(\lambda)\in \o_K$  and 
$F(\lambda)\overline{F(\lambda)}=(2N)^m$. 
Let $K = \QQ(\zeta_N)$ and $D$ be the decomposition group of $2$ in $K$
as before.
By hypothesis $($i$)$  we know that $[D:\QQ] = 2$.
While $p\equiv7\pmod8$ implies that $2$ splits in $\QQ(\sqrt{-p})$.
Hence $D =  \QQ(\sqrt{-p})$ is the decomposition group of $2$ in $K$.
We  now have the  following statement about the integer $m$.
\lemm{\label{lem_m=ordcle}
Let $E$ be any  extension of $D$ with $[E:D]\le2$
Let $\P$ be a prime in $E$ lying over $2$. Then 
the odd integer $m$ defined in  Theorem~$\ref{thm_pq}$ is equal to
the order of $\P$ in the class group $Cl(E)$.
}
\pf{
Let $\p=\P\cap D$ and denote by $a_p$ (resp. $b_p$) the 
order of $\p$ in $Cl(D)$ (resp. of $\P$ in $Cl(E)$).
By Gauss' genus theory  or class field theory \cite[Theorem $10.4($b$)$]{Washington},
 $Cl(D)$ is odd. So is $a_p$.
Remember  $D = \QQ(\sqrt{-p})$ and $2\o_D=\p\bar\p$ splits,
so $\p^{a_p}=(A+B\sqrt{-p})/2$ is principal in $\o_D$ for some integers $A$ and $B$.
It follows that $2^{a_p}\o_D = \p^{a_p}\bar\p^{a_p}=\frac{A^2+pB^2}{4}\o_D$
and thus $A^2+pB^2 =2^{a_p+2}$, which implies  $m\le a_p$ since $a_p$ is odd.

On the other hand, suppose  the equation $x^2+py^2=2^{m+2}$  has  integral solution $(A, B)$.
We know that both $A$ and $B$ should be odd.
Thus let $\delta=(A+B\sqrt{-p})/2\in\o_D$, 
we have $\delta\bar\delta\o_D=2^m\o_D=\p^m\bar\p^m$.
Since $m\le a_p$ and $a_p$ is the smallest integer such that
$\p^a_p$ is principal, we have $m=a_p$.

Now we observe that $Cl(D)\lr Cl(E)$ is injective by 
Proposition~\ref{prop_cl_inj}.
Thus $a_p=b_p$ and the result follows.
}
Let  $\P$ be as in Lemma $\ref{lem_m=ordcle}$.
Then $\vP:=\P\o_K$ is a prime ideal of $K$ since $D$ is the decomposition field.
Also $2\o_K=\vP\bar\vP$.
We use the following lemma to characterize  $F(\lambda)$. 
\lemm{ \label{lem_F_pq}
Suppose  $f$ is a GBF of  type  $[m, q=2p_1^{r_1}p_2^{r_2}]$ where $m, p_1, p_2, r_1, r_2$ is defined as in Theorem $\ref{thm_pq}$.
Then there exists an $o_K$-ideal $\vA$ not divisible by $\vP$ nor $\bar\vP$,
 such that for each $\lambda\in \ZZ ^m _q$,  
we have $F(\lambda)\o_K=\vA\vP^m$ or $\vA\bar\vP^m$.
}
\pf{
Since $f$ is a GBF of  type  $[m, q=2p_1^{r_1}p_2^{r_2}]$, we have \eq{ \label{thm_pq.eq_F_abs}
F(\lambda)\overline{F(\lambda)}=(2p_1^{r_1}p_2^{r_2})^m.
}
By hypothesis $($ii$)$  we known that 
$\ord_{p_2^{r_2}}(p_1)$ is odd and thus the complex conjugation is in
the decomposition group of $p_1$ in $K$. 
This implies that every prime in $K$ lying over $p_1$ is fixed by the complex conjugation.
We also observe that $\sqrt{-p_1}\in\o_K$ and 
$(\sqrt{-p_1})^2\o_K=p_1\o_K$.
By consider the order of each  prime in $K$ lying over $p_1$ appearing in    the equation \eqref{thm_pq.eq_F_abs},
we known that  $ F(\lambda)/ (\sqrt{-p_1})^{m r_1} \in \o_K$.
And similarly by hypothesis $($iii$)$,  $ F(\lambda)/ (\sqrt{p_2})^{m r_2} \in \o_K$.
So  if we set 
\eq{\label{eq_def_alpha}
\alpha=\frac{F(\lambda)}{ (\sqrt{-p_1})^{m r_1}(\sqrt{p_2})^{m r_2}},
}
then $\alpha\in\o_K$ and $\alpha\bar\alpha=2^m$.
Now we use Lemma \ref{lem_descent}  to obtain 
 $\beta\in\o_E$ and $\beta\bar\beta=2^m$  for some 
$\beta=\pm\zeta^j\alpha\in\o_E$ and $j\in\ZZ$,
where $E$ is a subfield of $K$ containing $D$ with  $[E:D]\le2$.

Recall  that $\ord_N(2)=\varphi(N)/2$  and $D\subseteq E$, so
$2\o_E=\P\bar\P$
  is the prime decomposition of $2$ in $E$.
Then we have $\beta\bar\beta\o_E= \P^m\bar\P^m$.
Note that $m$ is the order of $\P$ in $Cl(E)$ by Lemma \ref{lem_m=ordcle}, 
so we know $\beta\o_E=\P^m$ or $\bar\P^m$.
By \eqref{eq_def_alpha}  we have 
\eq{\label{thm_pq.eq_F_explicit}
F(\lambda) = \alpha (\sqrt{-p_1})^{m r_1}(\sqrt{p_2})^{m r_2}
= \pm\beta\zeta^j (\sqrt{-p_1})^{m r_1}(\sqrt{p_2})^{m r_2}.
}
Hence 
\eqn{
F(\lambda)\o_K =\P^m (\sqrt{-p_1})^{m r_1}(\sqrt{p_2})^{m r_2}\o_K
=\vA\vP^m\text{ (or }\vA\bar\vP^m\text{)},
}
where $\vA:= (\sqrt{-p_1})^{m r_1}(\sqrt{p_2})^{m r_2}\o_K$ is 
clearly not divisible by $\vP$ nor $\bar\vP$. So we complete the our proof.
}
\rk{\label{thm_pq.rk_new}
From \eqref{thm_pq.eq_F_explicit}  we can easily  verify that such
$F(\lambda)$ satisfy the equation 
\eqn{
 F(\lambda)\overline{F(\lambda)}= q^n.
}
So   the case  here is different from Feng's \cite{feng-gbf1,feng-gbf2,Feng3}
and hence the result stated in this theorem is new.
}
Now we turn to the 
\pf{[Proof of the Theorem $\ref{thm_pq}$]
Let $m, q$ be as hypotheses. Let $f: \ZZ_q^m\lr \ZZ_q$ be a function with Fourier transform $F(\lambda)$.
Our task is to  establish the property \eqref{eq_notin2} and \eqref{eq_same_class}
of $F(\lambda)$ needed in Proposition~$\ref{prop_epm}$.
Suppose  $\lambda \in \ZZ ^m _q$ and  $v\in \ZZ^m _q$  an element of order 2, 
we have $\zeta^{x\cdot v}_q=\pm1$, so
\eq{\label{thm_pq.eq_F+F}
F(\lambda)+F(\lambda+v)
=\sum_{x\in \ZZ^t_q}{\zeta_q ^{f(x)-x\cdot \lambda}(1+\zeta_q ^{-x\cdot v})}
\in 2\o_K.
}
Lemma~$\ref{lem_F_pq}$ tells us that 
 $F(\lambda)\o_K$ and  $F(\lambda+v)\o_K$ can only be one of the
two ideals $\vA\vP^m$ and $\vA\bar\vP^m$ 
 where $\vA$ is not divisible by $\vP$ nor $\bar\vP$.
In particular, $F(\lambda)\o_K\not\subseteq 2\o_K=\vP\bar\vP$ and $\eqref{eq_notin2}$
follows. 
If we   assume $F(\lambda)\o_K$  and $F(\lambda+v)\o_K$
 are different, say, $F(\lambda)\o_K = \vA\vP^m$ and $F(\lambda+v)\o_K=\vA\bar\vP^m$.
 Then  \eqref{thm_pq.eq_F+F} tells us  $F(\lambda+v)\o_K\in\vP$ since $2\o_K\in\vP$.
 This contradicts to the assumption that $F(\lambda+v)\o_K=\vA\bar\vP^m$.
 Hence $F(\lambda)\o_K=F(\lambda+v)\o_K=\vA\vP^m$. 
This proves \eqref{eq_same_class}. 

 By applying  Proposition~$\ref{prop_epm}$  we know that such function $f$ is
 not a GBF and we finish the proof.
}

\section{Non-existence result for GBFs of the type $[n,2p^e]$ for certain $n$ and $p$}\label{sec_sp}
We will prove Theorem \ref{thm_sp} in this section.
First we fix some additional notations. 
Suppose $n$ is odd, $N=p^e$ where $p\equiv7\pmod8$ is a prime.
Let  $K=\QQ(\zeta_N)$ and  $D\subset K$ be the decomposition field of $2$ in $K$ as before.
Since $\fracn{2}{p}{}=-1$,  $f=\ord_N(2)$ is odd. Thus $g:=\varphi(p^e)/f$ is even and we set $u=g/2$.
Suppose the  prime decomposition of $2$ in $D$ is
 \eq{\label{eq_2_dec}
2\o_D=\P_1\P_2\dots\P_g.
}  

If there are GBFs of type $[n,q=2N]$ then $F(\lambda)\overline{F(\lambda)}=(2N)^n$.
As in the proof of Lemma~\ref{lem_F_pq}, if we set 
$\alpha=\frac{F(\lambda)}{ (\sqrt{-p})^{em}}$ then 
 $\alpha\in\o_K$ and $\alpha\bar\alpha=2^n$.
Apply  Lemma \ref{lem_descent}  we  descent the above equation to a
 subfield $E\subseteq K$ with $[E:D]\le2$. Note $[E:D]$ divides 
the odd number $f$ it follows that $E=D$. Thus we obtain
 $\beta\in\o_D$ and $\beta\bar\beta=2^n$  for some 
$\beta=\pm\zeta^j\alpha\in\o_D$ and $j\in\ZZ$.

Since  $f$ is odd, the complex conjugation is not in the decomposition group of $2$.
Thus we may assume $\P_{u+k}=\bar\P_k,\ k=1,2,\dots,u$.
Then we have \eqn{
\beta\bar\beta\o_D=\prod_{k=1}^u\P_k^n\bar\P_k^n.
}
So \eq{\label{eq_beta_dec}
\beta\o_D=\prod_{k=1}^u\P_k^{n_k}\bar\P_k^{\bar n_k}
}
 where $n_k, \bar n_k$ are nonnegative integer such that 
$n_k+\bar n_k=n$ for all  $k=1,2,\dots,u$.

For convenience we write $x_k$ for $\P_k$ in $Cl(D)$ and view $Cl(D)$ additively. 
Hence \eqref{eq_beta_dec} becomes
\ga{
\sum_{k=1}^u(n_k x_k + \bar n_k \bar x_k) = 0  \label{eq_orig_rel} \\
\text{where }n_k+\bar n_k=n,\ k=1,2,\dots,u.
}
With the above notations, we  prove the following
\prop{\label{prop_sp_general}
Let $N=p^e$ where $e$ is a positive integer and $p\equiv7\pmod8$ is a prime.
Let  $n_0$ be the least odd integer such that \eqref{eq_orig_rel}
has  nonnegative integral solution 
$(n_1,n_2,\dots,n_g)$, where
$n_k+\bar n_k=n_0$ and  $n_{u+k}:=\bar n_k,\ k=1,2,\dots,u$.
If $n < n_0$ is a positive odd integer,
then there is no  GBFs of  type $[n ,2N]$.

Further more, if $n=n_0$ and  for all corresponding  solutions
$(n_1,n_2,\dots,n_g)$,
the set 
\eqn{Z_p(n_1,\dots,n_g):=\set{j = 1,2,\dots,g | n_j=0} }
 are nonempty and pairwise different,
then there is also no  GBFs of  type $[n_0 ,2N]$.
}
\pf{
The first assertion is trivially true 
since \eqref{eq_diophantine} is not solvable over $\o_K$   by the previous augment.
For the second one, we use the element partition method described in Section
\ref{sec_epm}.
The augment is similar to the one in the proof of Theorem~\ref{thm_pq}.
Suppose   $v$ is of order $2$, we have 
\eq{\label{thm_sp.eq_F+F}
F(\lambda)+F(\lambda+v)
=\sum_{x\in \ZZ^t_q}{\zeta_q ^{f(x)-x\cdot \lambda}(1+\zeta_q ^{-x\cdot v})}
\in 2\o_K.
}
We will show that  $F(\lambda)\o_K=F(\lambda+v)\o_K \not\subseteq 2\o_K$
and the proposition  follows from Proposition \ref{prop_epm}.
We know that 
\eqn{
F(\lambda)\o_K=\pm\zeta^j(\sqrt{-p})^{em}\prod_{k=1}^g\vP_k^{n_k}
} 
where $\vP_k:=\P_k\o_K$ is the prime in $K$ lying over $\P_k$
 and $(n_1,n_2,\dots,n_g)$  is a solution of \eqref{eq_orig_rel}.
The hypothesis that  \eqn{
Z_p(n_1,\dots,n_g)=\set{j = 1,2,\dots,g | n_j=0}
} is
nonempty  means  that
\eqn{
F(\lambda)\o_K\not\subseteq2\o_K=\vP_1\vP_2\dots\vP_g.
}
If we   assume $F(\lambda)\o_K$  and $F(\lambda+v)\o_K$
 are different, then
the corresponding solutions of \eqref{eq_orig_rel} are also different.
Since by hypothesis the corresponding $Z_p$'s are different,
we can assume that $\vP_j\mid  F(\lambda)\o_K$ and $\vP_j\nmid  F(\lambda+v)\o_K$
for some $j$. 
 Then  the decomposition $2\o_K=\vP_1\vP_2\dots\vP_g$ and 
 \eqref{thm_sp.eq_F+F} tell us  $F(\lambda+v)\o_K\in\vP_j$,
which  contradicts to the assumption that $\vP_j\nmid  F(\lambda+v)\o_K$.
 Hence $F(\lambda)\o_K=F(\lambda+v)\o_K$ and the proposition follows.
}

The above proposition is not concrete. To obtain the non-existence results in 
Theorem \ref{thm_sp} we  have to exploit the relations between $x_k$'s in $Cl(D)$.
By \eqref{eq_2_dec} we have
\eq{\label{eq_sum_x}
\sum_{k=1}^g x_k=0.
}
We want to find more relations.

Let $K_0=\QQ(\zeta_p)$ and $F=\QQ(\sqrt{-p})$ so 
we know that $F\subseteq K_0\subseteq K$.
Suppose further that
\eq{\label{eq_ordp_ordpe}
2^{p-1}\not\equiv1\pmod{p^2}.
}
One can check that this implies 
\eqn{
\frac{\varphi(p)}{\ord_p(2)}
= \frac{\varphi(p^e)}{\ord_{p^e}(2)}=g.
}
It follows that $D$ is also the decomposition group of $2$ in $K_0$.
Let  $K_0^+=K_0\cap\RR=\QQ(\zeta_p+\zeta_p^{-1})$.
Then  Miller's work on class number of $K_0^+$ gives
\thmu{[\cite{miller2015real}, Theorem $1.1$]
The class number of 
$\QQ(\zeta_p+\zeta_p^{-1})$ is $1$  if $p\le151$ is a prime.
}

From now on we  suppose   $p=31$ or $151$. 
Then $p\equiv7\pmod 8$ and one can check that the equation  \eqref{eq_ordp_ordpe}
is true.
Let $D^+=D\cap\RR\subseteq K_0^+$. Then $D$ is a quadratic extension
over $D^+$. 
By Miller's result and Lemma~\ref{lem_h_div}, we have
$h(D^+)=h(K_0^+)=1$.
Now   $2\o_{D^+}=\p_1\p_2\dots\p_u$ where $\p_k\o_D=\P_k\bar\P_k$,
 and all $\p_k$ is principal since  $h(D^+)=1$.
This implie  the relations
\eq{\label{eq_conj_x}
x_k+x_{u+k}=0,\quad k=1,2,\dots,u.
}

However, these relations above are not enough. We need  the
Stickelberger ideal introduced in Section~\ref{sec_facts}.
Let $\P=\P_1$ and correspondingly $x=x_1$.
Let $c$ be a integer not divisible by $p$. 
Since it is well-known that $p$ is the minimal integer such that $F\subseteq\QQ(\zeta_p)$,
it follows that $p$ is also the minimal one such that $D\subseteq\QQ(\zeta_p)$.
By  Proposition~\ref{prop_stick}, we have 
\eq{\label{eq_orig_ann}
(c-\s_c)\theta\ \P=1\text{ in }Cl(D).
}
Let $w$ be a primitive root mod $p$. Then 
$D=\left<2\right>=\left<w^g\right>\subseteq G_0=(\ZZ/p\ZZ)^\t$. It follows that we can assume
\eq{\label{eq_sigma_x}
\s_{w^{tg+s}}(x)=x_{s+1},\quad t\in\ZZ,\ s=0,1,\dots,g-1.
}
Let $k_{c,a}=[\frac{ca}{p}]$ for any integer $a$. We have \aln{
(c-\s_c)\theta&=(c-\s_c)\sum_{a\in(\ZZ/p\ZZ)^\t}\left\{\frac{a}{p}\right\} \s_a^{-1} \\
&=\sum_a \left(c\left\{\frac{a}{p}\right\}-\left\{\frac{ca}{p}\right\}\right)\s_a^{-1}\\
&=\sum_{a=1}^{p-1}k_{c,a}\s_a^{-1} &(\text{the definition of }k_{c,a})\\
&=\sum_{s=0}^{p-2}k_{c,w^{-s}}\s_w^s &(w^{-s}\text{ means } w^{-s} \mod p)\\
&=\sum_{t=0}^{f-1}\sum_{s=0}^{g-1}k_{c,w^{-(tg+s)}}\s_w^{tg+s}
}
Then by \eqref{eq_orig_ann} we have
\aln{
0&=\P^{\sum_{t=0}^{f-1}\sum_{s=0}^{g-1}k_{c,w^{-(tg+s)}}\s_w^{tg+s}}\\
&=\sum_{t=0}^{f-1}\sum_{s=0}^{g-1}k_{c,w^{-(tg+s)}}\s_w^{tg+s}(x)\\
&=\sum_{t=0}^{f-1}\sum_{s=0}^{g-1}k_{c,w^{-(tg+s)}}x_{s+1} &(\text{by \eqref{eq_sigma_x}})\\
&=\sum_{s=1}^g m_{c,s} \sum_{t=0}^{f-1}k_{c,w^{-tg+s-1}} x_s.
}
If we set \eq{\label{eq_m_cs}
m_{c,s}= \sum_{t=0}^{f-1}k_{c,w^{-tg+s-1}}
}
we have $p-1$ linear equations
\eqn{
\sum_{s=1}^g m_{c,s}x_s = 0,\quad c=1,2,\dots p-1.
}
We now combine these $p-1$ equations, together with the equation \eqref{eq_sum_x} and
the $u$ equations \eqref{eq_conj_x}, to give a whole   collection of equations \eq{\label{eq_ann}
XM^T=0} where $M$ is a $(p+u)\t g$ matrix with integer entries made of 
the coefficients of all  the $p+u$ equations and $X=(x_1, x_2,\dots, x_g)$.
To simplify these relations of $x_1,x_2, \dots,x_g$, we need to calculate  the 
\emph{Hermite normal form} of $M$. By the well-known result 
(c.f. \cite[\S~$2.4.2$]{Cohen}) for the existence of the Hermite normal form,
there exists a unique matrix  $U\in\GL_{p+u}(\ZZ)$, such that $H=M^TU$ is a Hermite normal form.
It follows from \eqref{eq_ann} that \eqn{
XH=0.}
In fact, $H$ can be obtained by applying  a finite sequence of elementary row operations over 
$\ZZ$ from $M^T$.

Now with the help of a computer and
using a simple program or a computer algebra system,
 we can calculate the individual Hermite normal form $H$ for
$p=31$ and $151$ ($u=3$ and $5$, respectively). Thus we obtain the relation

\eq{\label{eq_final_rel_31}
(x_1, x_2,x_3)\begin{pmatrix}
18&14&3\\
0&2&1\\
0&0&1\end{pmatrix}=0
}
and 
\eq{\label{eq_final_rel_151}
(x_1, x_2,\dots,x_5)\begin{pmatrix}
3934&1430&390&464&2457\\
0&2&0&0&1\\
0&0&2&0&1\\
0&0&0&2&1\\
0&0&0&0&1\end{pmatrix}=0,
}
where we omit $x_{u+1},\dots,x_g$ and other parts of $H$
since  $x_{u+k}=-x_k$.

Using these computational results, we can turn to
\pf{[Proof of the Theorem $\ref{thm_sp}$]
If $p=31$ the first column of the matrix in \eqref{eq_final_rel_31} tells us 
that $18x_1=0$ in $Cl(D)$. 
By \cite[Table~$\S3$]{Washington} we know $h_{31}^-$ is odd
and hence $h_{31}=h(K_0)=h_{31}^-h_{31}^+=h_{31}^-$ is also odd.
It follows that $9x_1=0$ and $\ord(x_1)=1,3$ or $9$.

We claim that $\ord(x_1)=9$.
Recall   $F=\QQ(\sqrt{-p})=\QQ(\sqrt{-31})\subseteq D$  and let $\p_F=\P_1\cap\o_F$.
It is easy to know that  $h(F)=3$ and  $\p_F$ has order $3$ in $Cl(F)$.
If $\ord(x_1)=1$, i.e. $\P_1=1$ in $Cl(D)$, then taking norm gives $\p_F=1$ in $Cl(F)$,
which is a contradiction.
If $\ord(x_1)=3$, then $\left<x_1\right>\cong\ZZ/3\ZZ$  and we may assume 
$x_1=1\mod3$. Then the second column of the matrix reads $14x_1+2x_2=0$,
or $x_2=-7x_1$. Hence $x_2=2x_1=-1\mod3$ and similarly $x_3=1\mod3$.
Thus $x_k=\pm1\mod3\in\ZZ/3\ZZ$ for all $k=1,2,\dots,6$.
But we know three of all six $x_k$'s (i.e. $\P_k$'s) lie over $\p_F$.
Suppose that $\p_F\o_D = \P_{k_1} \P_{k_2} \P_{k_3}$.
If all these three $x_{k_1},x_{k_2},x_{k_3}$ are the same, say $1\pmod3$, then 
since  $Cl(F)\lr Cl(D)$  is injective (Proposition~\ref{prop_cl_inj})
we have $\p_F=1$ in $Cl(F)$, a contradiction.
Otherwise we may assume $x_{k_1}=-x_{k_2}=1$ and then $x_{k_1}+x_{k_2} = 0$.
 Taking norm gives $\p_F^2=1$, which
is also false.

Thus we have $\ord(x_1)=9$ and using the matrix again we obtain
\eqn{
(x_1,x_2,\dots,x_6)=(1,2,4,-1,-2,-4)
}
 are all in $\left<x_1\right>\cong\ZZ/9\ZZ$.
We now apply Proposition \ref{prop_sp_general}.
Let $n=1,3\dots$ and solve the equation  \eqref{eq_orig_rel} modulo $9$.
A simple calculation tells us that $n_0=3$  in the proposition
and all the solutions
$(n_1,n_2,\dots,n_6)$ to \eqref{eq_orig_rel} corresponding to  $n=n_0$ are
\aln{
&(2,0,1,1,3,2),\ (2,2,0,1,1,3),\ (3,0,3,0,3,0),\ (3,2,2,0,1,1)\text{ and }\\
&(1,3,2,2,0,1),\ (1,1,3,2,2,0),\ (0,3,0,3,0,3),\ (0,1,1,3,2,2).
}
Obviously the corresponding $Z_p$'s
 are nonempty and pairwise different.
Hence we obtain by the proposition the non-existence of GBFs of type $[n\le3,2\t31^e]$.

The argument  for $p=151$ is similar. Using the matrix in \eqref{eq_final_rel_151} 
we know that $2\t7\t281x_1=0$. Noting that $h_{151}$ is also odd, we 
find that $\ord(x_1)=7,281$ or $1967$. 
In this case $F=\QQ(\sqrt{-157})$.
Knowing  that  $h(F)=7$ and  $\p_F$ has order $7$ in $Cl(F)$,
the candidate order $1$ and $7$ can be removed by the previous 
method. If we have $281x_1=0$, taking norm gives  $\p_F^{281}=1$,
which contradicts to $\ord(\p_F)=7$. Thus   $\ord(x_1)=1967$ and  we obtain
$x_1,\dots,x_{10}\in\left<x_1\right>\cong\ZZ/1967\ZZ$ and
\gan{
(x_1,x_2,\dots,x_5)=(1,-715,-195,-232,335)\\
x_{5+k}=-x_k,\ k=1,2,\dots,5.
}
Let $n=1,3\dots$ and solve the equation  \eqref{eq_orig_rel} modulo $1967$.
We find that $n_0=5$  in Proposition~\ref{prop_sp_general}
and all the solutions
$(n_1,n_2,\dots,n_6)$ to \eqref{eq_orig_rel} corresponding  to  $n=n_0$ are
\aln{
&(4,1,4,1,5,1,4,1,4,0)\ (5,3,2,5,5,0,2,3,0,0)\text{ and }\\
&(1,4,1,4,0,4,1,4,1,5)\ (0,2,3,0,0,5,3,2,5,5),
}
whose corresponding $Z_p$'s
 are nonempty and pairwise different.
Again  the proposition implies the non-existence of GBFs of type $[n\le5,2\t151^e]$.
The proof is done.
}

\section{Non-existence results  by PARI/GP and other remarks} \label{sec_pari}
If  $p\equiv1\pmod8$ and $\ord_{p^e}(2)$ is odd, 
there is no proof for non-existence of the type $[n,2p^e]$.
The Stickelberger relation method used in the previous section is 
not available since $K$ dos not contain a imaginary quadratic field and most $h_p^+$'s equal
to $1$ (Miller's conjectures).

However, if the degree of the decomposition field of $2$ is small,
we could use \url{PARI}\footnote{See \cite{pari}, a widely used computer algebra system 
designed for fast computation in number theory 
and originally developed by Henri Cohen and his co-workers}
to calculate the relations of $\P_k$'s in $Cl(D)$
(c.f. Section \ref{sec_sp}, especially \eqref{eq_2_dec}).
For example, we obtain without proof that 
\con{
Let $e$ and $n$ be  positive integers and $n$ be  odd.
Then  there is no GBFs of  type
$[1\le n\le7,2\t233^e]$.
}

The method combining Stickelberger relations and  Hermite
normal form is not a systematic one, 
so we hope that we could develop this method to obtain 
non-existence results for a class of types.

\newnoindbf{Acknowledgment}
The author would like to thank Yingpu Deng and Yupeng Jiang  for many helpful discussions and comments.

\bibliography{\bibfilename}
\bibliographystyle{amsplain} 
\end{document}